\documentclass{sigchi}
\usepackage{siunitx}
\usepackage[utf8]{inputenc}
\usepackage{float}
\usepackage{subfig}
\usepackage[show]{chato-notes} % Change the ``show'' to ``hide'' for submission to remove all \todo and \note's.
\usepackage{amsmath}
\usepackage[titlenumbered,ruled]{algorithm2e}
\usepackage{algpseudocode}
\usepackage{pifont}
\usepackage{times}
\usepackage{booktabs}
\usepackage{multirow}
\usepackage{balance}  % to better equalize the last page
\usepackage{graphics} % for EPS, load graphicx instead 
\usepackage[T1]{fontenc}
\usepackage{txfonts}
\usepackage{mathptmx}
\usepackage[pdftex]{hyperref}
\usepackage{color}
\usepackage{booktabs}
\usepackage{textcomp}
\usepackage{microtype} % Improved Tracking and Kerning
\usepackage{ccicons}  % Cite your images correctly!

\def\plaintitle{}

\def\emptyauthor{}
\def\plainkeywords{}

\makeatletter
\def\url@leostyle{%
  \@ifundefined{selectfont}{
    \def\UrlFont{\sf}
  }{
    \def\UrlFont{\small\bf\ttfamily}
  }}
\makeatother
\def\pprw{8.5in}
\def\pprh{11in}

\definecolor{linkColor}{RGB}{6,125,233}
\hypersetup{%
  pdftitle={\plaintitle},
  pdfauthor={\emptyauthor},
  pdfkeywords={\plainkeywords},
  bookmarksnumbered,
  pdfstartview={FitH},
  colorlinks,
  citecolor=black,
  filecolor=black,
  linkcolor=black,
  urlcolor=linkColor,
  breaklinks=true,
}

\begin{document}

\title{From Migration Corridors to Clusters:\\
The Value of Google+ Data for Migration Studies%
\titlenote{\textcolor{red}{\textbf{This is a pre-print of a paper accepted to appear at ASONAM'16.}}}}

\numberofauthors{3}
\author{%
\alignauthor
Johnnatan Messias\\
       \affaddr{Universidade Federal de Minas Gerais}\\
       \affaddr{Belo Horizonte, Brazil}\\
       \email{johnnatan@dcc.ufmg.br}
\alignauthor
Fabricio Benevenuto\\
       \affaddr{Universidade Federal de Minas Gerais}\\
       \affaddr{Belo Horizonte, Brazil}\\
       \email{fabricio@dcc.ufmg.br}
\alignauthor
Ingmar Weber\\
       \affaddr{Qatar Computing Research Institute}\\
       \affaddr{Doha, Qatar}\\
       \email{iweber@qf.org.qa}
\alignauthor
Emilio Zagheni\\
       \affaddr{University of Washington}\\
       \affaddr{Seattle, USA}\\
       \email{emilioz@uw.edu}
}

\maketitle

\begin{abstract}
%Recently, there has been a considerable amount of efforts that explore online data, especially from social networks, to investigate migration patterns across countries. Previous efforts showed that online data is not only valuable to identify the main migration corridors and the reasons behind these migration flows, but is also easy to obtain in comparison with traditional offline data based on high costly surveys. Existing efforts are mainly focused on exploring migration between a pair of countries. Despite the extreme importance of these efforts, little is known about the migration causes and patterns when we look into groups of countries. In this work, we use data extracted from Google+ to give a first step in this direction.

Recently, there have been considerable efforts to use online data to investigate international migration. These efforts show that Web data are valuable for estimating migration rates and are relatively easy to obtain. However, existing studies have only investigated flows of people along migration corridors, i.e.\ between \emph{pairs of countries}. In our work, we use data about ``places lived'' from millions of Google+ users in order to study migration `clusters', i.e.\ groups of countries in which individuals have lived  sequentially. For the first time, we consider information about more than two countries people have lived in. We argue that these data are very valuable because this type of information is not available in traditional demographic sources which record country-to-country migration flows independent of each other. We show that migration clusters of country triads cannot be identified using information about bilateral flows alone. 
% look at potential reasons for migrants to sequentially move to a set of countries. To do that 
To demonstrate the additional insights that can be gained by using data about migration clusters, we first develop a model that tries to predict the prevalence of a given triad using only data about its constituent pairs. We then inspect the groups of three countries which are more or less prominent, compared to what we would expect based on bilateral flows alone.  Next, we identify a set of features such as a shared language or colonial ties that explain which triple of country pairs are more or less likely to be clustered when looking at country triples. Then we select and contrast a few cases of clusters that provide some qualitative information about what our data set shows. 
The type of data that we use is potentially available for a number of social media services. We hope that this first study about migration clusters will stimulate the use of Web data for the development of new theories of international migration that could not be  tested appropriately before.

%\todo{complete here}

\end{abstract}

\section{Introduction} \label{sec:intro}

%\todo[john]{Change the csc concepts}

Advances in our understanding of demographic processes have historically been the result of a graceful dance between new theories and new data. In some areas of demographic research, e.g., the study of mortality and fertility, large-scale data collections that include censuses, vital registration systems, and surveys have profoundly enhanced our knowledge of population dynamics. On the other hand, concerning migration studies, lack of data and issues related to cross-country harmonization of existing sources have drastically limited our ability to test theories \cite{de2010overcoming, Laczko2015}.

Web data have features that are qualitatively different from existing traditional sources and that can be leveraged to evaluate migration theories and their predictive power. In this article, we present a study of migration systems that relies on Google+ data. More specifically, we analyze the extent to which the frequency of people who have lived in three distinct countries is related to bilateral migration flows for pairs of countries. We particularly focus on country triads that occur more or less often than expected given only the data for pairwise flows. The analysis that we present in this article is only possible because our data set of places where Google+ users have lived allows us to evaluate the relative frequencies of triadic groups of countries in which users have lived. This type of information is typically not available in traditional demographic sources which only track movement between pairs of countries.

International migration systems are clusters of countries that are characterized by large exchanges of people and by related feedback mechanisms that connect the countries in terms of flows of goods, capital, information, and ideas. These systems typically persist over time \cite{Massey1993}. One mainstream empirical approach for identifying migration systems is to assess changes over time in bilateral flows of migrants for all countries \cite{zlotnik1992empirical,dewaard2012migration}. This approach is problematic partly because reliable data on bilateral flows for a large number of countries, and over time, are not available. In addition, ``the trouble with this approach is that the system becomes little more than a summary of flows.'' \cite{bakewell2013relaunching}

We argue that lack of data constrains the definition of migration systems to a summary of flows. However, with better data, such as self-reported ``places lived'' that are typically available for a number of social media sources,  we can deepen our understanding of migration systems. With the additional knowledge of migration clusters, individual migration corridors are no longer observed independently, yielding a higher level knowledge of migration patterns.

To illustrate that bilateral migration flows (expressed as pairs of countries in which people have lived) are not sufficient to predict more complex migration clusters (triads of countries in which people have lived), Table~\ref{tab:toy_example} provides a simplified example. In the hypothetical situation there are two scenarios, each with four migrants. Both scenarios generate the same distribution of bilateral flows, each occurring exactly once. But they differ in the migration clusters that are observed. Similarly, other scenarios can easily be constructed where either all possible clusters or no cluster at all are observed while, again, the distribution of bilateral migration flows is identical.

\begin{table}[ht]
\centering
\begin{tabular}{cc|cccc|l} &    & \multicolumn{4}{c|}{Countries Lived In} & \hspace{7mm}Bilateral Flows\\
&    & A & B & C & D & \\ \hline
\multirow{3}{*}{\rotatebox{90}{Scenario 1}} & M1 & x & x & x &  & (A,B), (A,C), (B,C)  \\
& M2 & x &   &   & x & (A,D) \\
& M3 &   & x &   & x & (B,D) \\
& M4 &   &   & x & x & (C,D) \\ \hline \hline
\multirow{3}{*}{\rotatebox{90}{Scenario 2}} & M1 &   & x & x & x & (B,C), (B,D), (C,D) \\
& M2 & x & x &   &   & (A,B) \\
& M3 & x &   & x &   & (A,C) \\
& M4 & x &   &   & x & (A,D) \\ \hline
\end{tabular}\caption{Two toy scenarios for four countries and four migrants illustrating that observing migration corridors is not sufficient to study migration clusters. In both cases, each of the six possible migration corridors is observed exactly once. However, the first scenario features the migration cluster (A,B,C) whereas the second features (B,C,D).}\label{tab:toy_example}
\end{table}

In this paper, we contribute to the literature about migration systems and show how new Web data can be used in the context of classic theories of migration. At the same time, the opportunities opened up by new data and Web science are likely to stimulate the development of new theories that could not be appropriately tested before.

This article is organized as follows. In Section~\ref{sec:related} we provide a review of the relevant literature. Section~\ref{sec:dataset} describes the data set of Google+ users that we analyzed. Section~\ref{sec:expected_migration_flows} presents our baseline model to estimate triadic groups of countries from bilateral flows. Section~\ref{sec:outlier_analysis} discusses those triads in which the frequency of people who have lived in all three countries is substantially higher or lower than what we would expect based on bilateral flows. The last section summarizes our results and offers some concluding remarks.

\section{Related Work} \label{sec:related}

%\todo{Include more related papers}

The study of human migration relies on accurate and up-to-date information that is often not available. Traditional demographic sources include censuses, population registers and sample surveys. These data have been extremely useful for improving our understanding of migration processes. However, they are far from perfect. Reliable migration statistics,  in particular estimates of flows of migrants, are not directly available for a number of countries. Thus these quantities are often estimated indirectly. For example, Abel and Sanders developed an approach to estimate the minimum sizes of international bilateral flows that are consistent with available estimates of stocks of foreign-born people~\cite{abel2014quantifying}.

The recent availability of geo-located Web data has stimulated the development of new approaches to study international migration. For example, Zagheni and Weber \cite{Zagheni:2012:YYE:2380718.2380764} and State \textit{et al.}~\cite{State2013} estimated international migration rates using IP-geolocated data of millions of anonymized Yahoo users' logins. These studies showed that it is feasible to estimate international migration rates, at a large scale, from logins to a website. They also pointed to important challenges related to the fact that the sample is not representative of the underlying population, and offered methodological contributions to deal with selection bias~\cite{Zagheni:2012:YYE:2380718.2380764,nikolaos2015demographic}.

%that   instance, t   They considered users whom, for a period of one year, spent three months in a different country of origin (migrants) or less than a month (tourists). The prediction of migration and tourism flows used as attributes are colonial ties, geographical location, economic development and visa control. Their analyzes showed the persistence of traditional migration patterns, such as the emergence of new routes. Migration tend to be more commuting between countries with borders. They also observed particularly pendular high levels within the European economic space even considering territorial restrictions.

Zagheni \textit{et al.}~\cite{Zagheni2014} and Hawelka \textit{et al.}~\cite{hawelka2014geo} have used geo-located Twitter tweets data to estimate international migration rates and trends. 
%from 500,000 \textit{Twitter} users. They considered only users belonging to the \textit{OECD}\footnote{Organisation for Economic Co-operation and Development} member countries.
They showed that estimates of international mobility rates are consistent with statistics about tourism \cite{hawelka2014geo}. When no official statistics are available for calibration, an approach based on the `difference-in-differences' technique proved useful to reduce bias in the data and to estimate trends~\cite{Zagheni2014}. Moreover, Twitter  geo-located data have a lot of potential for helping us understand the relationship between internal and international migration.

%From a methodological point of view, they showed that They analyzed a subsample of users who have posted \textit {tweets} regularly with geolocation data. Proposed \textit{difference-in-differences} approach to reduce the selection bias when trend in emigration rates for individual countries. Their results show that the approach is relevant to address two issues in migration literature: (1) methods can be used to predict points of \textit{turning} migration trend; (2) the ``Twitter"  geolocation data can improve an understanding of the relationship between internal and international migration.

State \textit{et al.}~\cite{State2014} looked into LinkedIn data to investigate trends in international labor migration. They estimated changes in residence, over time, for millions of users, based on their educational and professional histories reported on the LinkedIn website. They found that, conditional on being an international migrant with college education, the probability of choosing the United States as the destination decreased during the period from 2000 to 2012. This is partially related to the rise of migration corridors in East Asia and the dot-com bubble, as well as the great recession in the United States.

Recently, Kikas \textit{et al.}~\cite{kikas2015explaining} used data from the voice and video call service Skype to study international migration and its relationship to social network features. They found that the percentage of international calls, the percentage of international links and foreign logins in a country, together with information about GDP, could be used to produce relatively accurate proxies of migration rates.

Network theory has been widely used to explain international migration \cite{Massey1993}. The main idea is that interpersonal ties that link people in origin and destination countries reduce the costs and risks of migration and increase the expected returns to migration. The network theory of migration is very powerful. However, the lack of comprehensive data about social network connections among countries limit our ability to test and refine theories that explain migrations in terms of networks. 

In this paper we contribute to this area by looking at a previously untapped type of data source. We consider the countries people have lived in. This information can only be obtained from data on migration histories, which are typically not available in sample surveys. When some data exist, they are usually collected only for small regions of a country. Data about countries in which people have lived are potentially available for a number of social media services. To our knowledge, nobody has used this type of information to contribute to our understanding of international migration in the context of networks. We thus hope that our paper may stimulate more research in this area.

%Network of Skype call

%\todo{Write our approach summarization here...}

%Further details will be explained in the following sections.

\section{Google+ Dataset} \label{sec:dataset}

%This section describes the dataset we have used. It also describes the Google+ data extracted from user profiles that we have used. After obtained all features from the user profiles we had to filter these data as described in Data Improvements. Finally, we pointed the dataset limitations.

%\subsection{Google+ Data}
We used the dataset of all Google+ profiles that was collected by Magno et al.\ \cite{Magno2014} between March 23 and June 1, 2012. %When inspecting the site map 193,661,503 user IDs were found.
 The data set contains information for 160,304,954 Google+ profiles. % because some IDs have been deleted or were not able to extract their information.

For this article we focus on data about international migration. More specifically,  we extract the  Google+ field  ``places lived'' (``Places where I lived"). In this field, users list places in the world where they have lived. The items in the list are free text which means that (i) different languages are used (``United States'' vs.\ ``Estados Unidos''), (ii) different variations are used within the same language (``São Paulo" vs.\ ``Sampa"), and (iii) locations of different geographic granularities occur  (``Brazil" vs.\ ``Minas Gerais" vs.\ ``Belo Horizonte").
  Google+ automatically performs geo-coding and maps the free text entries to co-ordinates on Google Maps. For our study, we  used these co-ordinates and mapped them to countries. In total, our sample includes 22,578,898 (14\%) users with a geo-mapped location.

 \begin{figure}[!ht]
   \begin{center}
     \includegraphics[width=86mm]{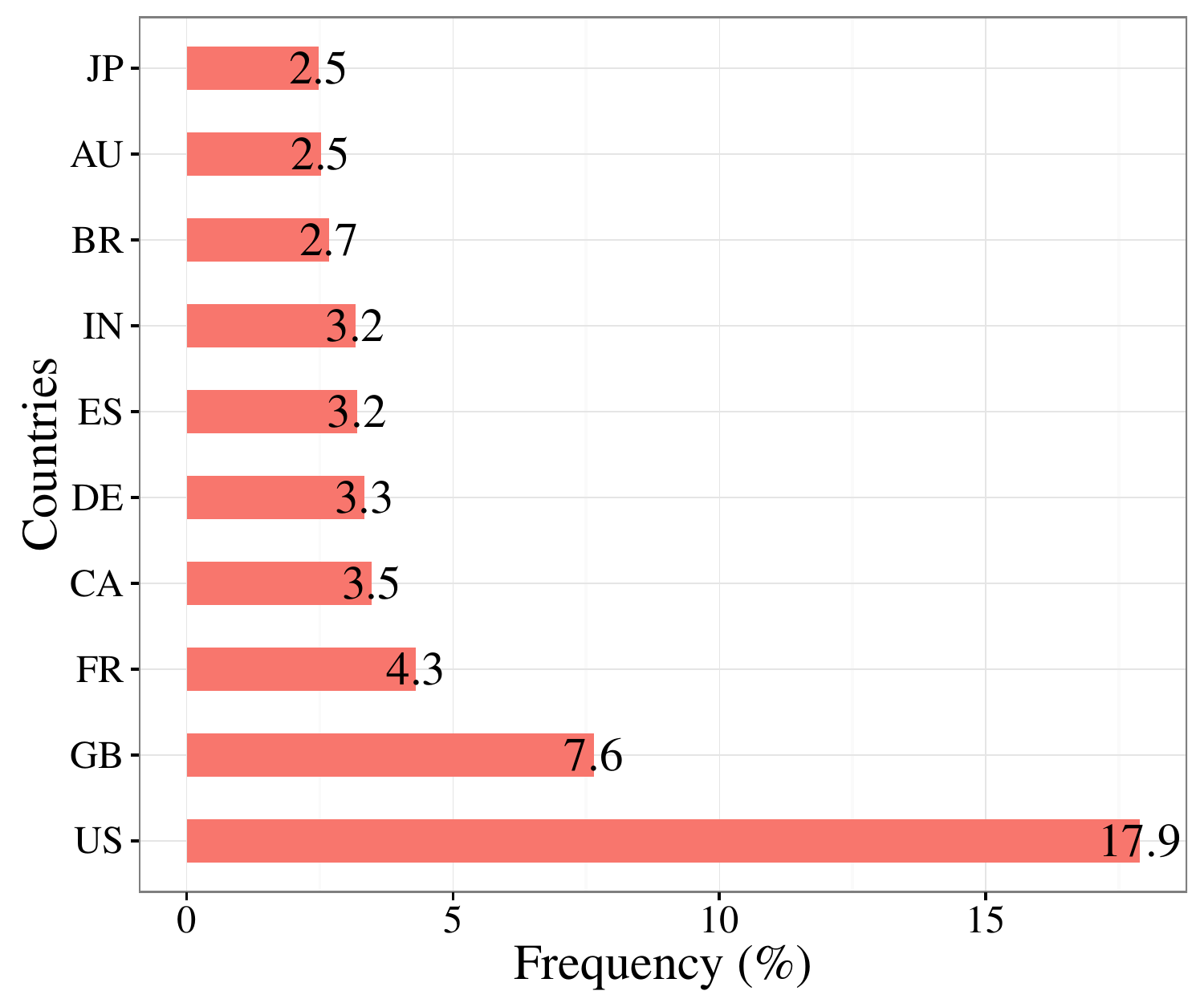}
   \end{center}
   \caption{Fraction of top 10 countries, in terms of number of users, in our data set.}
   \label{fig:top10fraction}
 \end{figure}

%In addition to the user countries we also analyzed the language of their posts. To identify the language of the posts we used the \texttt{langid.py}~\footnote{\url{https://github.com/saffsd/langid.py}}, a solution for identifying languages which provides the probability of a given text belongs to a certain language, working well for both long documents and for short documents, including microblogs~\cite{paper_langid}.

The ``places lived'', unfortunately, do not come in chronological order, e.g., either the first or the last location might indicate the user's country of origin. It is therefore impossible to tell if a user who lived both in the US and in India moved from India to the US or the other way around. Though this is an obvious limitation, our main analysis is centered around \emph{sets of countries where subsets of users have lived in}. In particular, we look at users who have lived in triples of countries (A,B,C) without distinguishing their order.

%\note[Ingmar]{@Johnnatan: do we ever use the number of users per country in our models? Does this number include the single-country users?}

As our study is about international migration, we only considered the subset of users who have lived (``places lived'') in at least two distinct countries. We refer to this group of users as \emph{migrants}.
% of our sample that is  do not consider data for the XXX\% \todo[Ingmar]{@Johnnatan: how many out of 22.6M?} of users who have ``places lived'' from a single country. Only users with at least two distinct countries are considered.
%\note[Ingmar]{@Johnnatan: Please confirm that all these percentages are with respect to the 1.96M migrant users.}
Our dataset includes 1,958,656 migrants. Users who lived in USA correspond to 17.9\% of the data set, while for GB the fraction is 7.6\% (see Figure~\ref{fig:top10fraction}). In terms of the number of distinct countries users have lived in, (i) 1,565,803 have two countries in their list, (ii) 271,142 have three, (iii) 69,129 have four countries, and (iv) 52,582 have at least five.
%
%\note[Ingmar]{@Johnnatan: Just to confirm: we did not normalize and so a user with many countries will generate more pairs, right? (That's fine, just to know.)}
%

In order to avoid data sparsity issues for countries with very few migrant users, we only considered countries that have at least 1,000 people who have lived there. There are 192 such countries.

For each migrant user, we extracted all the pairs and triples of valid countries they lived in. For example, if a user has lived in countries \{BR, FR, HU\}, then we would generate the set of country pairs \{(BR, FR), (BR, HU), (FR, HU)\} as well as the triple (BR, FR, HU). Countries in pairs and triples are listed in alphabetical order to have a canonical form, but no chronological order is implied. For each pair and triple we count how often it occurs among our 1.96M migrant users. In the following, we will also refer to country pairs found in our data as ``migration corridors'', and to country triples as ``migration clusters''.
Our analysis looks at how the counts for the migration corridors relate to the corresponding clusters. In particular, we are interested in finding and explaining counts for migration clusters which are unusually high or low, given the counts of the contributing migration corridors.

Aiming at allowing reproducibility we release our migration dataset to the research community. The dataset is available at \url{http://www.dcc.ufmg.br/~fabricio/migration-dataset/}.

\section{Expected Migration Flows}\label{sec:expected_migration_flows}

Our data set enables us to identify clusters of countries that are connected through people who have lived in all of them at some point. We then assess whether the frequency of particular clusters in our data set is higher or lower than what we would expect purely on the basis of frequencies of pairwise connections between countries (number of users who have lived in two countries).  For example, if we observe certain migration flows among the pairs of countries (UK, USA), (India, USA), and (India, UK), respectively, intuitively one could expect that the number of Google+ users that lived in the cluster (India, UK, USA) is somehow proportional to these bilateral flows. We want to investigate just how strong this proportionality is and, in particular, which factors are linked to over- or under-proportionate counts of particular migration clusters.  In other words, our general goal is to identify and study cases where observed counts of people who have lived in three countries are higher or lower than expected. By `expected', we mean the counts that one would predict if one only knew data for bilateral migration flows, i.e., pairs of countries in which users lived in.

Here we present our  approach to define the expected migration flow of a cluster. For simplicity, we only consider cluster sizes of three countries. However, our methodology easily generalizes to larger cluster sizes, though data sparsity quickly becomes a limiting factor for tuples of more than three countries.
We formulate the comparison of ``more or less than expected'' as a ranking comparison task. Concretely, we rank clusters both (i) according to a function associated to the pairwise counts and (ii) according to their actual frequencies in our Google+ data. The relative difference in the positions between the predicted and observed rankings is then our measure of interest.

Note that the functional dependency between the pair and triple counts is not a priori clear and would depend heavily on assumptions of how migrants move. As we are interested in \emph{discovering} such patterns, we try to avoid overly specific modeling assumptions and, instead, experiment with four different formulas to see which gives the best match between the predicted and observed rankings. All these four formulas merely (i) are symmetric in the three edges, i.e., there is no ``first'' or ``second'' edge, and (ii) their predicted frequency of triples increases with increases in the individual pairwise counts.

\begin{itemize}
  \item $\mbox{Ranking 1} \sim freqAB + freqAC + freqBC $\\
  \item $\mbox{Ranking 2} \sim freqAB * freqAC * freqBC $\\
  \item $\mbox{Ranking 3} \sim \min(freqAB, freqAC, freqBC) $\\
  \item $\mbox{Ranking 4} \sim \min(freqAB,freqAC,freqBC) * \\  mean(freqAB,freqAC,freqBC) $\\
\end{itemize}

% \begin{equation}
% \begin{aligned}
% \mbox{Ranking 1} \sim freqAB + freqAC + freqBC\\
% \end{aligned}
% \end{equation}

% \begin{equation}
% \mbox{Ranking 2} \sim freqAB * freqAC * freqBC\\		
% \end{equation}

% \begin{equation}
% \mbox{Ranking 3} \sim \min(freqAB, freqAC, freqBC)\\
% %\mbox{Ranking 3} \sim \frac{freqAB + freqAC + freqBC}{log10(freqAB * freqBC * freqAC)}\\
% \end{equation}

% %\todo[Emilio]{@Johnnatan:make sure that the equation for Ranking 4 is spaced across two rows or that it fits within the space}

% %\begin{equation}
% \begin{multline}
% %\mbox{Ranking 4} \sim \frac{freqAB * freqAC * freqBC}{log10(freqAB + freqBC + freqAC)}\\	
% %\mbox{Ranking 4} \sim  {\min(freqAB,freqAC,freqBC) * \\ mean(freqAB + freqAC + freqBC)}\\
% Ranking 4 \sim  \min(freqAB,freqAC,freqBC) \\ * mean(freqAB,freqAC,freqBC)\\
% \end{multline}

%\end{equation}

\noindent
where $freqAB$, $freqAC$, $freqBC$ are the frequencies of migrations flows among the three pairs of countries of a cluster (A, B, C).

Intuitively, as the observed summed counts of the pairs in a triangle increase, the corresponding observed triple counts should also increase. This is why we included  (freqAB + freqAC + freqBC) in our baseline `Ranking 1'. The model `Ranking 2' is inspired by approaches to the study of migration flows known as gravity models \cite{cohen2008international}. These models explain flows between origin and destination countries as proportional to the product of their sizes and inversely proportional to their distances. Here we consider that the effect of distance on triples of countries where users lived in is implicitly accounted for by the number of users who have lived in the respective pairs of countries. `Ranking 2' is appealing because it is intimately connected to a class of models, gravity models, that have been used quite successfully by migration scholars.  For our specific situation, however, it is also clear that the \emph{minimum} value of the three pairwise counts plays an important role as, trivially, the triple count is upper bounded by the minimum of the three pairwise counts. In other words, when we consider a system of three countries, the maximum number of people who have lived in all three countries cannot be larger than the minimum value of the number of people who have lived in only two of the three countries.  To take this dependency into account, we also included  $\min(freqAB,freqAC,freqBC)$ in our baseline `Ranking 3'. The model `Ranking 4' is a further extension that adds to `Ranking 3' by including the average size of the pairwise frequencies. The intuition is that the larger the migration system, the higher the probability that people who have lived in two countries might have been attracted to a third country as well.

In order to measure the extent to which these rankings produce accurate results, we compare them with the ground truth data from Google+.
Table~\ref{tab:ranks} shows the correlation of these rankings with the ground truth ranking according to two well-known rank correlation measures: Kendall and Spearman rank correlation coefficients~\cite{abdi2007kendall}.
We can see that Ranking 4 yields the best prediction of the actually observed Google+ cluster ranking, using only information from pairs of countries.  In the rest of the paper we refer to this ranking as the \emph{expected ranking}.

\begin{table}[!htb]
    \caption{Performance of ranking formulas}
    \label{tab:ranks}
    \centering
   \begin{tabular}{| l | l | l | }
  \hline
  Description		&	Kendall				&	Spearman			\\
  \hline
	Ranking 1		&	0.350			&	0.498				\\\hline
	Ranking 2		&	0.546			&	0.737			\\ \hline
	Ranking 3		&	0.502			&	0.689			\\\hline
\textbf{Ranking 4}	&	\textbf{0.565}	&	\textbf{0.754}		\\
  \hline
  \end{tabular}
\end{table}

%\todo[fabricio]{We need a motivating example that explore the expected ranking. I already add some text here just make clear what I mean.}

The creation of an expected ranking from pairs of countries enables us to gain some insights about how countries are integrated in terms of people who have lived in all of them. For example, in our data set, 1,077 people have lived in Great Britain (GB), Malaysia (MY), and Singapore (SG). This number, freq(GB,MY,SG), is substantially larger than what we would expect from the counts of users who have lived in two of these countries: freq(GB,MY)=5,552; freq(GB,SG)=6,642; freq(MY,SG)=7,242. This means that within this group of countries, users who have lived in two of them have a relatively high probability to have lived in the third country.   In this situation, the observed value for the cluster is \emph{higher than expected}. Conversely, when we consider the cluster formed by Great Britain (GB), the Philippines (PH), and the United States (US), we observe that a similar number of users (1,022) have lived in all the three countries. However the pairwise frequencies are substantially higher:  freq(GB,PH)=3,179; freq(GB,US)=152,976; freq(PH,US)=24,599. In this case a large number of users have lived either in the Great Britain and the US, or in the Philippines and the US. However, only a small proportion of these users have lived in all the countries. The observed number of users who have lived in the three countries is lower than what we expected based on pairwise frequencies.  We refer to this situation as  \emph{lower than expected}.

In the next section we formulate a classification problem where we investigate the discriminative power of
additional features, such as a shared language, colonial link, distance, to differentiate clusters.
%Here, we are particularly interested in the discriminative power of different features to distinguish between the classes.

%Instead of going over all the cases on an individual basis, we formulate a classification problem where we use additional features, such as a shared language, to \emph{predict} which clusters will have higher-than-expected counts. Here, we are particularly interested in the discriminative power of different features to distinguish between the classes.

\if 0

\begin{table}[!htb]
    \caption{Rankings metrics}
    \label{tab:ranks}
    \small
    \centering
   \begin{tabular}{| l | l | l | l |}
  \hline
  Description		&	Kendall				&	Spearman			&	$R^2$	\\
  \hline
	Ranking 1		&	0.328971			&	0.469137			&	0.2200892	\\
\textbf{Ranking 2}	&	\textbf{0.553483}	&	\textbf{0.744599}	&	\textbf{0.5544277}	\\
	Ranking 3		&	0.350253			&	0.497591			&	0.2475965	\\
	Ranking 4		&	0.546198			&	0.736994			&	0.5431608	\\
  \hline
  \hline
  \end{tabular}
\end{table}

\fi

\section{Explaining Deviance from \\ Expectation}\label{sec:outlier_analysis}

Our next step is about identifying a set of features related to migration clusters. The aim is to investigate their relative discriminatory power to distinguish clusters that are ranked higher than, lower than, or as expected.
First, we present a definition for three classes.

\subsection{Classes of Clusters}

%Figure~\ref{fig:cdf_diff} shows the cumulative fraction for the differences between the ground truth ranking position and the expected ranking. Positive values mean that the observed Google+ triple counts are higher than expected, whereas negative values mean the opposite. We can see a higher slope around 0, indicating that these differences follow a distribution that is similar to a normal distribution. Based on this observation, we divided our data into a 5-fold set of groups, each containing 20\% of the data. 
We rank the triples by how much their actual frequency ranking differs from the expected one. We then divide this ranking into five strata, each containing 20\% of the data. Based on this division, we consider the following three cluster classes. 

\begin{itemize}
\item \textbf{As expected}: We consider as expected or close-to-expected the center 20\% of the clusters with the expected and actual ranks approximately equal.
\item \textbf{Higher than expected}: We consider as higher-than-expected those clusters that appear in the top 20\% on the positive side.
\item \textbf{Lower than expected}: We consider as lower-than-expected those clusters that appear in the top 20\% on the negative side.
\end{itemize}

Thus, our approach neglects 40\% of the data, which corresponds to the folds that appear in between these three cluster classes we considered. For the observations that we do not consider, there is much more uncertainty associated to potential differences in ranking. 

%\todo[Ingmar]{@Johnnatan: Just to understand: the middle class contains 40\% of all clusters whereas the other two each contain 20\%, correct? And 20\% are ignored.}

%\todo[fabricio]{Johnnatan. Please, fix this Figure~\ref{fig:cdf_diff}. Remove the gray in the background, use large font, Y axis should be 100 as you say percentage or just say CDF and let the y-axis from 0 to 1.}

%\begin{figure}[!htb]
%  \centering
%    \includegraphics[width=0.49\textwidth]{images/cdf_diff}
%  \caption{Cumulative distribution function (CDF) of the differences between the ground truth and the expected ranking}
%  \label{fig:cdf_diff}
%\end{figure}

\subsection{Features}
%\todo[fabricio]{Describe Features here}

Migration patterns depend on a multitude of factors. The goal of our analysis is to understand which type of features (derived from the triads), e.g., geographical or historical, either lead to or inhibit the formation of migration clusters. This type of analysis is impossible with traditional data sources which only record pairwise migrations independently.% We accessed the dataset from~\cite{10.1371/journal.pone.0122543} to investigate the number of countries in each cluster that are part of a common civilization, common region, GDP, common colonial link, common language, and visa requirement. To do that, we operationalized it as a single numeric score, with values from 0 to 3, that represent the number of pairs in the triad of countries for each feature.

%Next, we briefly describe the features we consider to help us understanding why some clusters obtain such a high or low ranking position than expected. 
%These features were obtained by the State et al.~\cite{10.1371/journal.pone.0122543}

\begin{itemize}

\item \textbf{Common Civilization}: A recent study~\cite{10.1371/journal.pone.0122543} has found empirical evidences, from online data, that  
eight culturally differentiated civilizations can be identified, as theoretically posited by Huntington~\cite{huntington1997clash}, with the divisions corresponding to differences in language, religion, economic development, and spatial distance. %In short, these civilizations are Sinic, Hindu, Islamic, Latin American, Western, Orthodox, African, and Buddhist.   
%We accessed their dataset to investigate the number of countries in each cluster that are part of a common civilization.
We operationalized it as a single numeric score, with values 0, 2, or 3, that represent the number of countries (None, 2 out of 3, and All) in the triad of countries with common civilization. The same approach of assigning a single integer to a triple was used for Common Colonial Link, Common Language, and Visa Requirement. 
%The data set from~\cite{10.1371/journal.pone.0122543} also contains information about the features we considered next.  

\item \textbf{Geographic Distance}: The distance among countries represents a physical barrier for migration.
For each cluster we consider as features the average distance among the pairs, as well as the maximum and minimal distances between the pairs of countries within the cluster.  The distances were obtained from the geolocation\footnote{\url{http://opengeocode.org/download/cow.txt}} (latitude, longitude) of the center of the mass of each country. Thus, the distance between countries is calculated by the spherical distance, considering the earth curvature.
Another geographic related feature is the common region, which represents the main continental regions in which countries are grouped. %For this specific feature, we operationalize the score as a single numeric feature, with values from 0 to 3, that represent the number of pairs of countries within the same common region.

\item \textbf{GDP}: The gross domestic product (GDP) is one of the primary indicators used to gauge the size of a country's economy. It represents the total dollar value of all goods and services produced over a specific time period. 
    For each cluster we consider as features the average GDP among the pairs, as well as the maximum and minimum GDP between a pair of countries within the cluster.

\item \textbf{Common Colonial Link}: This feature aims at capturing if two countries share a 
colonial past. %Similarly to Common Civilization and common region, we consider it as  a numeric feature, with values from 0 to 3, that represents the number of pairs of countries with common colonial link.

\item \textbf{Common Language}: This feature aims at assessing if two countries share the same language. %It  consists of  a numeric value (from 0 to 3) that represents the number of pairs of countries with common colonial link.

\item \textbf{Visa Requirement}: Visa requirement may represent another barrier for migration. %We consider it as  a numeric feature, with values from 0 to 3, that represent the number of pairs of countries that require visa.

\end{itemize}

\begin{figure}[!htb]
  \centering
    \includegraphics[width=0.49\textwidth]{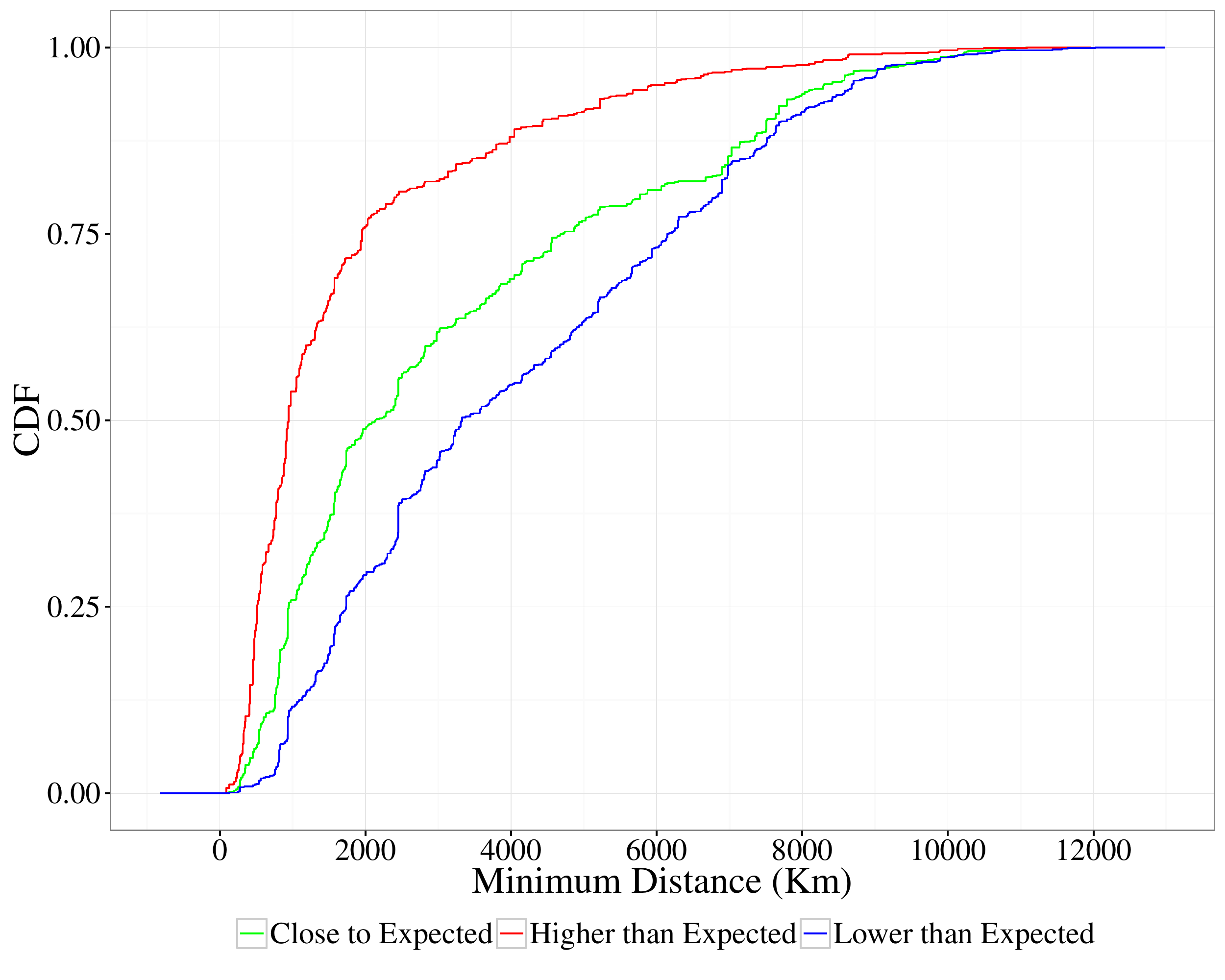}
  \caption{Cumulative distribution function (CDF) for the feature minimum distance for the three cluster classes}
  \label{fig:cdf_min_dist}
\end{figure}

\begin{figure}[!htb]
  \centering
    \includegraphics[width=0.49\textwidth]{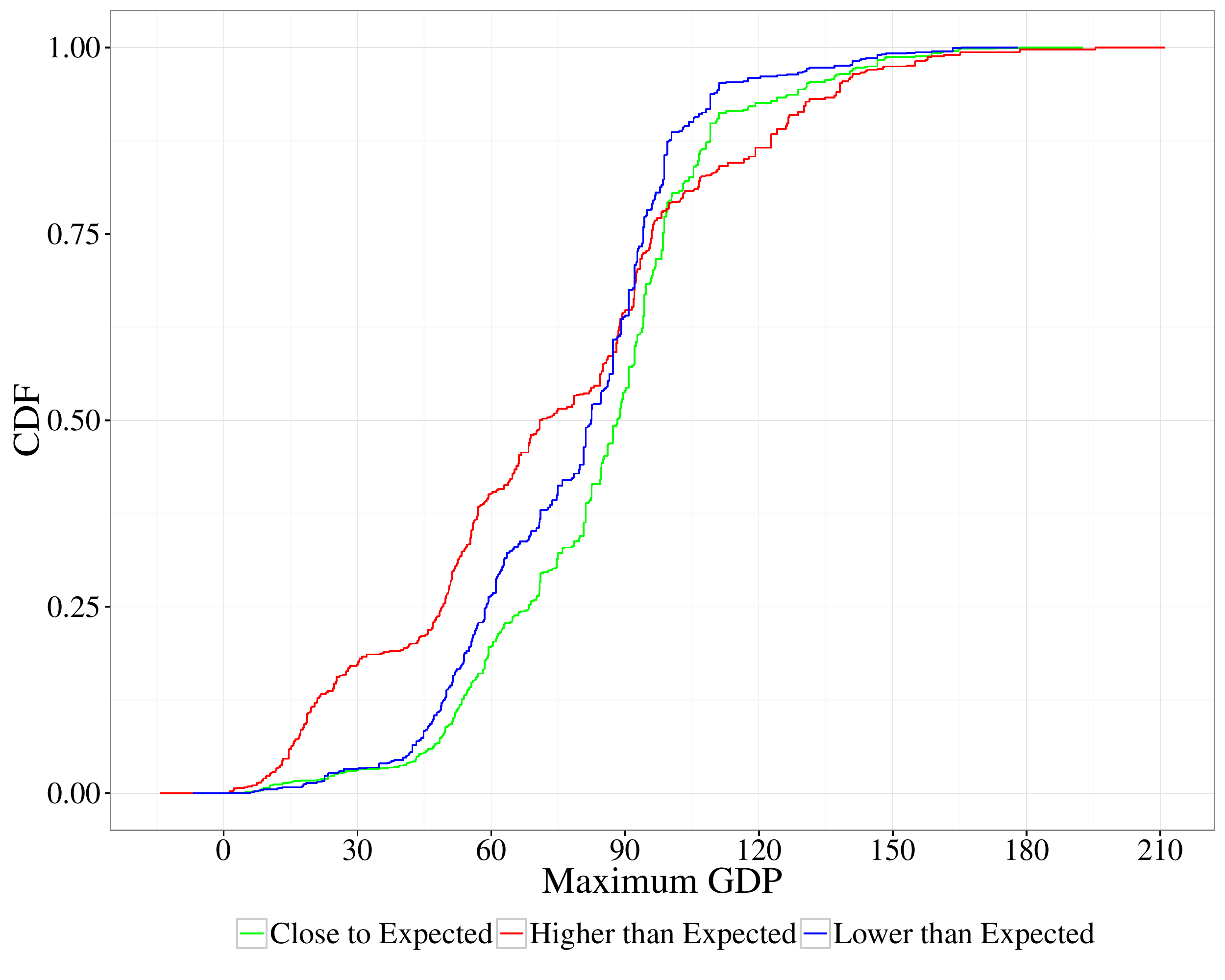}
  \caption{Cumulative distribution function (CDF) for the feature maximum GDP for the three cluster classes. GDP values are expressed in hundreds of billions of USD}
  \label{fig:cdf_max_gdp}
\end{figure}

Figure~\ref{fig:cdf_min_dist} and Figure~\ref{fig:cdf_max_gdp} show the cumulative distribution function for features \textit{minimum distance} and \textit{maximum GDP} for the three cluster classes, respectively. We can note that 75\% of the pairs of countries within the cluster higher-than-expected are within 2,000 Km in distance, whereas only around 27\% of the pair of countries within the cluster lower-than-expect are within this same distance. Similarly, we can note that 50\% of the pairs of countries within the cluster close-to-expected have GDP lower than 88 (hundreds of billions of USD), a higher value in comparison with the other cluster classes (49\% for higher-to-expected and 82\% for lower-than-expected).

\begin{figure*}[!htb]
  \centering
    \includegraphics[scale=0.5]{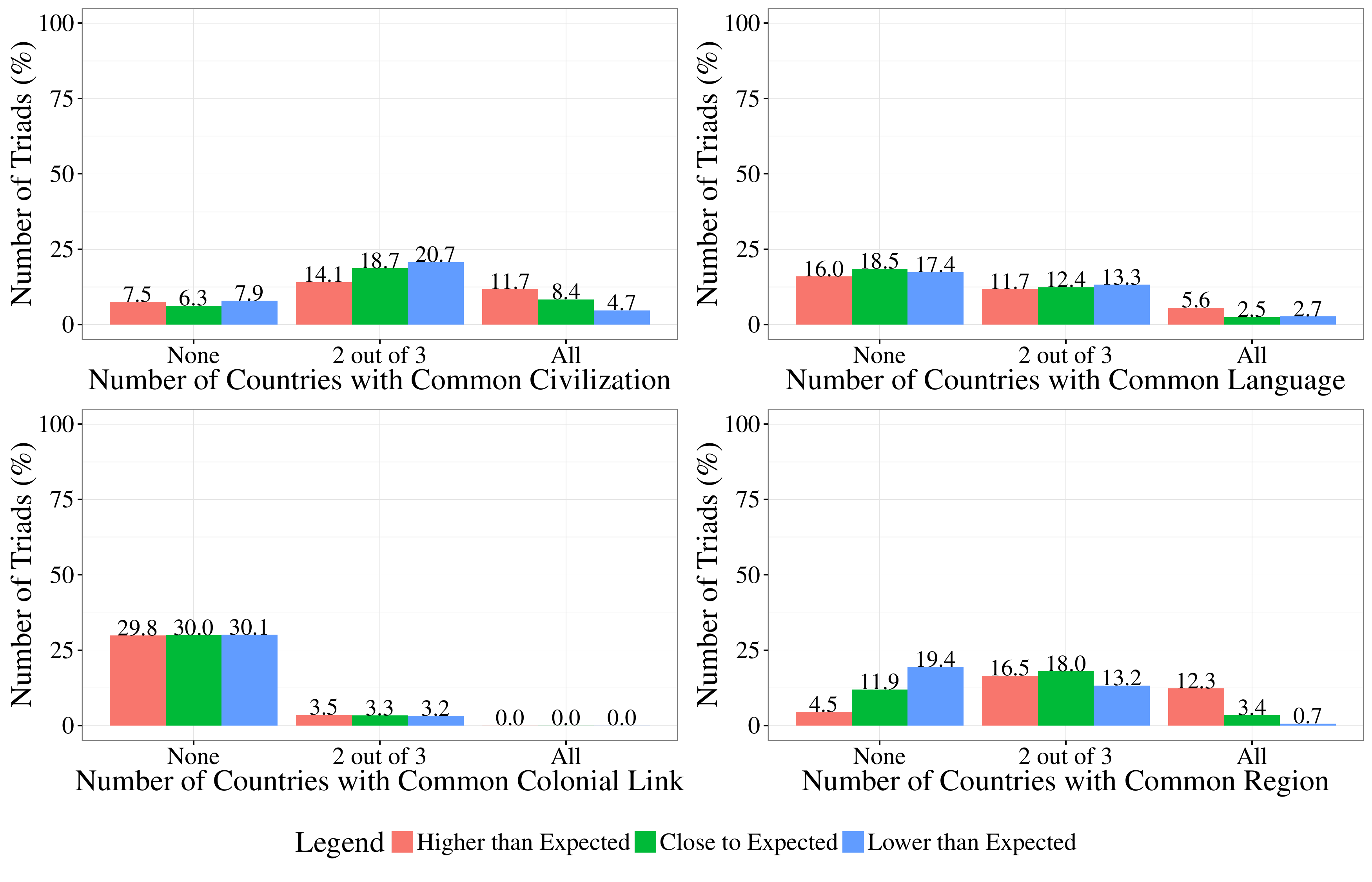}
  \caption{Distribution of the difference between the ground truth and the expected ranking considering four features that account common factors among  countries}
  \label{fig:common_factors}
\end{figure*}

Figure~\ref{fig:common_factors} shows the difference between the ground truth and the expected ranking considering four features that account for common factors among countries. Particularly, we show the amount of countries (out of 3, because of the triad) within each cluster class with common civilization, common language, common colonial link, and common region. We can see interesting trends here. For example, we can note triads in the cluster of higher than expected tend to have more countries with common civilization than the rest. We can also note a similar trend for common region and common language. On the other hand, colonial link shows a very similar distribution for all three classes. In the next section we provide a rank for these features in terms of their discriminative power to distinguish among classes.

%\note[fabricio]{I need a help with explaining these features. Do not know exactly the references and if Require visa is actually do not require visa!}

\subsection{Assessing Feature Importance}

We assessed the relative power of the features considered in discriminating one cluster class from the others by independently applying two well-known feature selection methods, namely, information gain and $\chi^2$ (Chi Squared)~\cite{feature2}. Table~\ref{tab:igxrank} shows the ranking of the most important features for differentiating the three classes 
(higher-than-expected, close-to-expected, lower-than-expected). We note that the four geographic distance features appear on the top of the table, followed by all the features related to GDP. %Similarly, when we look into the ranking of the most important features with discriminatory power to distinguish between only two cluster classes, i.e. higher-than-expected and lower-than-expected, as shown in Table~\ref{tab:igxrank_higher_lower_than-expected}; higher-than-expected and close-to-expected, as shown in Table~\ref{tab:igxrank_higher_than-expected}; and close-to-expected, lower-than-expected, as shown in Table~\ref{tab:igxrank_lower_than_expected}, 

Though the observation that geographic vicinity leads to migration clusters seems obvious, it is worth pointing out that it is not. As the geographic vicinity already increases the pairwise migration counts, it is implicitly already accounted for in the expected ranking of migration clusters. So what is observed here is a ``supra-linear'' type of effect that is not predicted by the pairs alone.

%We note that the top features are also related to geographic distance and GDP. This suggests that distance and GDP may represent the main driving forces among those we investigate that are able to differentiate clusters that are ranked higher than expected from lower than expected and from those that are considered close to expected.

\begin{table}[!htb]
    \caption{Ranking of most important features for differentiating the three classes (higher-than-expected, close-to-expected, and lower-than-expected), presented by the IG (Information Gain) Ranking and the ${\chi}^2$ (Chi-Squared) Ranking.}
    \label{tab:igxrank}
    \small
    \centering
   \begin{tabular}{| l | l | l | l | l |}
  \hline
  Description				&	IG Rank	& 	IG Value  	& ${\chi}^2$ Rank & ${\chi}^2$ Value		 \\
  \hline
Min Distance & 1 &0.231 & 1& 984.742 \\
Max Distance & 2 &0.180  &3& 767.547 \\
Common Region &3 &0.178  &2& 780.458 \\
Avg Distance  &4 &0.173 & 4& 745.858 \\
Max GDP &5 &0.102 & 5& 474.392 \\
Avg GDP &6 &0.089  &6& 408.225 \\
Min GDP &7 &0.070  &7& 312.460 \\
Common Civ. &8& 0.033 & 8 &147.838 \\
Common Visa &9 &0.017  &9& 80.004 \\
Com. Col. Link  &10 & 0.0001 & 10 & 0.679 \\
  \hline
  \end{tabular}
\end{table}

\subsection{Illustrative Cases}

In the previous section we attempted to summarize, in a quantitative way, the key features that discriminate various classes of countries according to our definition. Here we  discuss some examples that offer a more qualitative understanding of what we observed in the data. More specifically, we present a couple of cases in which the observed number of people who have lived in all three countries is higher than what we would have expected based on pairs of flows. We will then discuss a couple of cases for which the opposite is true.

Consider the United Arab Emirates, India  and Singapore. In our dataset, 805 users have lived in all the three countries. 17,584  users have lived in the United Arab Emirates and India. 7,665 users have lived in India and Singapore. A lower number of users, 1,970, have lived in the United Arab Emirates and Singapore. Based on pairs of flows, we would expect that a relatively low number of users have lived in all three countries. In fact our original ranking model 4 would rank this triple at place 682. However, in our Google+ dataset the actual ranking is number 200. About 40\% of the users who have lived in Singapore and in the United Arab Emirates have also lived in India. This indicates that in addition to the large communities of Indians in Singapore and in the United Arab Emirates, there is also a sizable unexpected community of users who have been in all the three countries and who strengthen interpersonal networks across these countries. 

Similarly, when we consider the cluster Spain, France, and Italy, we would expect to observe less people who have been in all three countries than what we actually find in the data. 2,322 users have lived in all the three countries; 15,455 have lived in Spain and France; 11,230 have lived in France and Italy; 9,628 have lived in Spain and Italy. Based on the flows for pairs of countries, our ranking model would have expected the triple to rank number 111, when in fact it ranked number 36 in our data set. This example might be related to the context of European integration that lowers the cost of moving to countries within the Union. Moreover, these countries are close in terms of distance, with languages that are relatively similar. In addition, interpersonal networks may be strong enough to make the cost of moving across these countries relatively low. Overall, we observe that  a substantial fraction (more than expected) of the people who have lived in two of these countries, have also lived in the third one.

The situation is quite different for the cluster composed of Brazil, Mexico, and the US. In our Google+ dataset, 14,593 users have lived in Brazil and Mexico; 46,784 users have lived in Brazil and the US; 67,065 users have lived in Mexico and the US. Although these pairs of flows are quite substantial, only 1,386 users have reported living in all the three countries. Brazil, Mexico, and the US have strong bilateral connections, but they do not seem to be integrated within a larger cluster in a demographic sense, meaning that people typically migrate only along one of the corridors. Our ranking model would have expected this triple to rank number 12 based on bilateral flows. Instead it ranked number 80 in the actual Google+ data.

Canada, China, and Great Britain offer a similar example of a weaker-than-expected cluster.  6,093 users have lived in Canada and China; 25,696 users have lived in Canada and Great Britain; 8,189 users have lived in China and Great Britain. However, only 623 users have lived in all the three countries. As for the previous example, migration does occur along the corridors but rarely within the whole cluster. For example, a number of Chinese students might go to study to Canada or Great Britain. However, only a relatively small fraction would experience living in both Canada and Great Britain. This example is important because it also highlights one of the limitations of our approach: Google+ is not accessible in China. Thus the  values that we observe for this cluster might be skewed, particularly towards Chinese living abroad, or non-Chinese people who have lived in China at some point.

%\subsection{Case Studies}
%\todo[Ingmar]{@Emilio: Can you analyze some examples ``by hand'' and give some conjectures? We should also have a map of some kind.}

\section{Conclusions and Discussion} \label{sec:conclusion}

We started this paper by saying that new theories and new data move hand-in-hand to advance our understanding of demographic processes. In this article, we showed that new data about `places lived' can lead to the development of new theories of international migration. We started with the observation  that data about `places lived' for more than two countries (migration histories) are traditionally not available, except for some special subregions within a particular country. This type of information is not equivalent to data about bilateral flows, and is very valuable to identify specific characteristics of high level migration systems. In particular, studies on what leads users to migrate within clusters of countries cannot be performed with data limited to pairwise migration flows.

We believe that this line of research is relevant and timely, and that the increasing availability of information about pseudo-migration histories from online sources opens new and exciting opportunities at the intersection of social network analysis and demography. Here we would like to discuss some of the limitations of our current research and point to some directions for future work.  

For this study, we work with a sample of Google+ users that is quite large and that can be collected at low cost. However, Google+ data have several shortcomings. First, as mentioned earlier, we do not know the chronological \emph{order} in which people have lived in the various countries that they list.  For our specific application, this is not a  problem since we are interested in how people connect countries by living in several of them.  However, more elaborate analyses could be performed if we could identify each user's home country and the countries of residence in a chronological order. This type of information has been used to evaluate bilateral flows of professional migrants on LinkedIn \cite{State2014}. The same type of dataset could be used to evaluate clusters of countries in terms of professional skills and the direction of flows within a cluster (for example, are people more likely to move from country A to country C via an intermediate step in country B?). 

Second, the Google+ dataset that we are using is neither representative of the world population nor of any specific country. Several different types of selection bias mechanisms affect our data. Users in our dataset are, first of all, Internet users. They are more likely to be more highly educated and younger than the average population, especially in the context of developing countries with low Internet penetration rates. As a result  our users are most likely more internationally minded and mobile than in the underlying populations. In fact, 9\%, 1.96M out of 22.6M users with at least one geo-coded location, are migrants in our dataset. This is substantially higher than the United Nations estimate of the percentage of people who live in a country different from their country of birth, which is between 3\% and 4\%. In addition, most of the Google+ users are located in North America or in Western Europe. The extent of bias differs from country to country. China is an extreme case, since the country is blocking access to Google and other popular social media services \cite{Bamman12censorshipand}. In our study we did not attempt to calibrate our results in order to remove the bias, as discussed in other venues \cite{nikolaos2015demographic}. Instead, we attempted to control for a number of biases by evaluating the number of people who have lived in three countries conditional on having information about bilateral flows. For example, since Google+ is quite popular in the US, we would expect more people in our data set to have lived in the US and in a second country. Conditional on having lived in these two countries, we considered the fraction of users who have lived in a third one and compared it with the expected value based on the size of bilateral flows. This is an imperfect correction that was appropriate for our specific application, but not necessarily generalizable to other situations. More research to address issues related to selection bias in social media data is certainly needed.

Third, there is a range of data quality issues. These include the free text nature of the ``places lived'' field, which could lead to ambiguities. In addition, we need to be aware of  potential misreporting or intentionally fabricated  histories.

In the end, no single dataset is enough to study international migration. In the future, we hope to be able to combine several data sources that include both Web data and traditional demographic sources. We hope that this paper contributes to highlight  the potential and weaknesses of Web data for the study of migration processes and that it would stimulate collaborations between researchers in the area of demography and Web science. 

Aiming at allowing reproducibility we release our migration dataset to the research community. The dataset is available at \url{http://www.dcc.ufmg.br/~fabricio/migration-dataset/}.

%As future work we plan to incorporate more features and assess their predictive power and also investigate to what extent these features can \emph{predict} which clusters will have higher-than-expected or lower-than-expected counts.

\section*{Acknowledgment} \label{sec:acknowledgments}

\small{This work was partially supported by the project FAPEMIG-PRONEX-MASWeb, Models, Algorithms and Systems for the Web, process number APQ-01400-14, and by individual grants from CNPq, CAPES, and Fapemig. We also would like to thank Gabriel Magno for sharing his data collection.}
\bibliographystyle{SIGCHI-Reference-Format}
\bibliography{references}

\end{document}